\documentclass[aps,prb,twocolumn,superscriptaddress,showpacs]{revtex4-1}
\usepackage{amssymb,graphicx,xspace,color}
\usepackage[pdfauthor={Frank Steckel}, bookmarks=true, pdfborder={0 0 0}, colorlinks=true,citecolor=blue,linkcolor=black,urlcolor=blue]{hyperref}
\newcommand{\tc}{$T_c$\xspace}

\newcommand{\ts}{$T_S$\xspace}

\newcommand{\nf}{NaFeAs\xspace}
\newcommand{\rnf}{NaFe$_{1-x}$Rh$_x$As\xspace}
\newcommand{\cnf}{NaFe$_{1-x}$Co$_x$As\xspace}
\newcommand{\cnfy}{NaFe$_{1-y}$Co$_y$As\xspace}

\begin{document}
\title{Combined resistivity and Hall effect study on NaFe$_{1-x}$Rh$_x$As single crystals}

\author{Frank Steckel}
\email[]{f.steckel@ifw-dresden.de}
\author{Robert Beck}
\affiliation{Leibniz-Institute for Solid State and Materials Research, IFW-Dresden, 01069 Dresden, Germany}
\author{Maria Roslova}
\affiliation{Department of Chemistry and Food Chemistry, TU Dresden, 01062 Dresden, Germany}
\author{Dirk Bombor}
\affiliation{Leibniz-Institute for Solid State and Materials Research, IFW-Dresden, 01069 Dresden, Germany}
\author{Igor Morozov}
\affiliation{Leibniz-Institute for Solid State and Materials Research, IFW-Dresden, 01069 Dresden, Germany}
\affiliation{Department of Chemistry, Lomonosov Moscow State University, 119991 Moscow, Russia}
\author{Sabine Wurmehl}
\affiliation{Leibniz-Institute for Solid State and Materials Research, IFW-Dresden, 01069 Dresden, Germany}
\affiliation{Institut f\"ur Festk\"orperphysik, TU Dresden, 01069 Dresden, Germany}
\author{Bernd B\"uchner}
\affiliation{Leibniz-Institute for Solid State and Materials Research, IFW-Dresden, 01069 Dresden, Germany}
\affiliation{Institut f\"ur Festk\"orperphysik, TU Dresden, 01069 Dresden, Germany}
\affiliation{Center for Transport and Devices, Technische Universit\"at Dresden, 01069 Dresden, Germany}
\author{Christian Hess}
\email[]{c.hess@ifw-dresden.de}
\affiliation{Leibniz-Institute for Solid State and Materials Research, IFW-Dresden, 01069 Dresden, Germany}
\affiliation{Center for Transport and Devices, Technische Universit\"at Dresden, 01069 Dresden, Germany}
\date{\today}
\begin{abstract}
Electrical transport measurements are used to study the Rh-doped NaFeAs superconductor series with a focus on the tetragonal phase. The resistivity curvature has an anomalous temperature dependence evidencing in the phase diagram two crossover regions of changes in the scattering rate, the effective mass as well as of the charge carrier density. The first crossover region is directly connected to the structural transition and resembles the onset of resistivity anisotropy. The second crossover region can as well be deduced from the temperature dependent Hall coefficient. A comparison to literature NMR data suggests this region to be connected with nematic fluctuations far above the tetragonal to orthorhombic phase transition. 
\end{abstract}

\maketitle

\section{Introduction}
Most of the FeAs-based superconductors share the same principles of their electronic phase diagram, i.e. i) an antiferromagnetic order following a structural transition in the undoped compounds ii) a suppression of these phases upon chemical doping or pressure, and iii) a dome-like behavior of a superconducting phase. The superconducting critical temperature \tc is the highest at the instance when complete suppression of the structural and magnetic phase is reached. Recent phase diagram studies show, that essentially the doped charge seems to influence the phase diagram. Thus, Co and Rh-doping in NaFeAs \cite{Steckel2015} and in BaFe$_2$As$_2$~\cite{Ni2009} as well as Ni and Pd doping comparably affect the transition temperatures. The phase transitions in the undoped and underdoped compounds came recently into focus because the rotational symmetry breaking seems to be triggered from the electronic system in the Fe-based superconductors. \cite{Fernandes2014, Boehmer2015} This transition is called nematic \cite{Kivelson1998} and happens naturally in twins such that only microscopic probes can locally detect the lowered two-fold symmetry in the Fe-plane of these materials. \cite{Rosenthal2014, Chuang2010} By applying a small strain to the crystal lattice an easy axis for the electronic distortion is defined and, thus, the material becomes detwinned. In this case even macroscopic methods can probe the difference in the orthogonal nematic $a$- and $b$-directions. For example in resistivity measurements of detwinned crystals a large anisotropiy between $\rho_a$ and $\rho_b$ is observed in many different Fe-based superconductors. \cite{Blomberg2013, Deng2015, Chu2010,Fisher2011,Jiang2013, Tanatar2010, Ying2011, Ma2014} It turned out, that already far above the structural transition temperature \ts such an anisotropy is measurable if uniaxial strain is applied. However, the strain field smears the transition\cite{Ren2015} and enhances the fluctuation regime. Thus, in order to study the zero-strain fluctuations other methods were applied. Nuclear magnetic resonance (NMR), \cite{Ma2014, Dioguardi2013, Ning2010} magnetic torque measurements, \cite{Kasahara2012, Xu2014} X-ray absorption spectroscopy,\cite{Kim2013} point contact spectroscopy \cite{Arham2013} as well as angle resolved photoemission spectroscopy (ARPES) \cite{Yi2011} were able to detect a fluctuation regime in doped 111 and 122 compounds. However, the question about the temperature and doping evolution of the fluctuation regime in the phase diagram of the Fe-based superconductors remains open. Therefore, a method highly sensitive to subtle fluctuations of the incipient transition is needed. 

The transport coefficients are capable of probing even tiny changes of the electronic structure and thus should be suited to detect fluctuations in the electronic system. The electrical resistivity has been proven powerful for detecting and analyzing similarly subtle electronic structure changes. For example, in La-doped Bi$_2$Sr$_2$CuO$_{6+\delta}$ and Sr-doped La$_2$CuO$_4$ \cite{Takagi1992, Ando2004} as well as in F-doped LaFeAsO and SmFeAsO \cite{Hess2009} the analysis of the resistivity slope and curvature allowed to detect a pseudogap-regime as well as the crossover from non-Fermi-liquid to Fermi-liquid behavior. Intimately connected with the resistivity is the Hall coefficient and is, thus, a natural candidate to cross-check such subtle electronic structure changes. \cite{Yan2013}

In this paper, we report a detailed analysis of the resistivity curvature and the Hall coefficient of Rh-doped \nf single crystals. Our results clearly show a crossover region intimately connected to $T_S$ and furthermore another crossover at very high temperatures traceable through the whole accessible electronic phase diagram. We show that the first region tracks the onset of the resistivity anisotropy whereas the second region evidences the incipient electronic fluctuations.

\section{Experimental}

Crystal growth and characterization of the Na$_{1-\delta}$Fe$_{1-x}$Rh$_x$As single crystals with $x=0-0.043$ is elaborated in detail in Ref.~\onlinecite{Steckel2015}. Due to the high sensitivity to air of Na$_{1-\delta}$Fe$_{1-x}$Rh$_x$As, all preparations and subsequent transport measurements have been done either in inert gas atmosphere (Argon) or in vacuum. 

The crystals were contacted with a two component silver epoxy in the standard four point contact geometry inside an argon box and afterwards securely closed inside a homemade probe rod. The evacuated probe rod had then been inserted to a Helium bath cryostat. The resistivity measurements have been already presented in Ref.~\onlinecite{Steckel2015} to determine the phase transition temperatures. The Hall effect measurements were conducted with magnetic fields up to $\pm15$~T. The perpendicular resistivity $\rho_{xy}$ has been antisymmetrized with respect to the magnetic field.

\section{Results}
\subsection{Resistivity}
\begin{figure}
\includegraphics[clip,width=1\columnwidth]{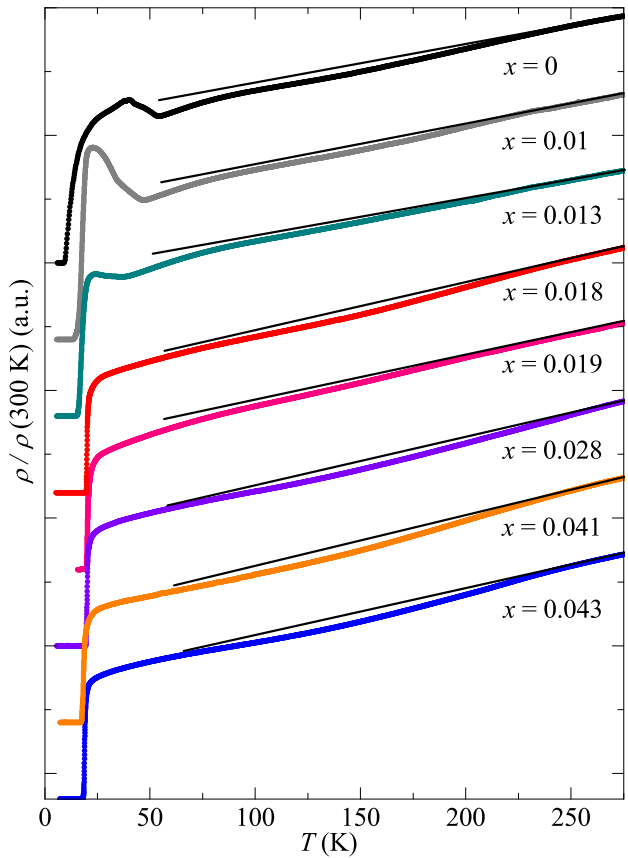}
\caption{(Color online) Normalized and shifted resistivity of the \rnf single crystals from Ref.~\onlinecite{Steckel2015}. Together with the resistivity a linear fit to the high temperature region is shown as a black solid line. Thus, the deviation from this linear behavior as a broad dip in the intermediate temperature regime becomes visible.}
\label{fig:1}
\end{figure}

Fig.~\ref{fig:1} displays the resistivity of the \rnf single crystals. At high temperatures $T>250$~K a linear fit to the resistivity data is possible and extrapolated to lower temperatures. Already here a deviation from the linear extrapolation is visible in the temperature range below $\sim 225$~K. Resulting from the canonical picture of a metal for $T>\Theta_D/4\sim 75$~K (with $\Theta_D$ the Debye-temperature of NaFeAs\cite{Presniakov2013,Wang2012}) the electron-phonon-scattering rate is expected to be linear in temperature (cf. the Bloch-Gr\"uneisen formula\cite{Ziman2007}) and thus with zero curvature of $\rho(T)$. Any deviation from this behavior at high temperatures points directly to unusual scattering or additional changes in the charge carrier density $n$ or their effective mass $m$.

All crystals show a strong deviation from the canonical linear behavior independent of the doping level. The deviation has its maximum in the intermediate temperature regime of approximately 150~K. Interestingly, this maximal deviation does not shift with increasing Rh content and, thus, seems to be unaffected from the suppression of the structural and magnetic phase. Upon lowering the temperature further, the deviation from the linear extrapolation becomes smaller. Below temperatures of $\sim 50$~K the temperature dependence of the resistivity is dominated by the phase transitions of the structural and magnetic ordering and superconductivity yielding typical hump and dip anomalies. In our analysis, we therefore focus on the temperature-range $\gtrsim50$~K and below 175~K in the tetragonal phase. In Ref.~[\onlinecite{Spyrison2012}] similar deviations have been reported for Co-doped \nf and are argued as having a notion towards a change of the effective mass and charge carrier density. Nevertheless, all these effects can naturally be ascribed to changes of the scattering rate, too.

\begin{figure}[!t]
\includegraphics[clip,width=1\columnwidth]{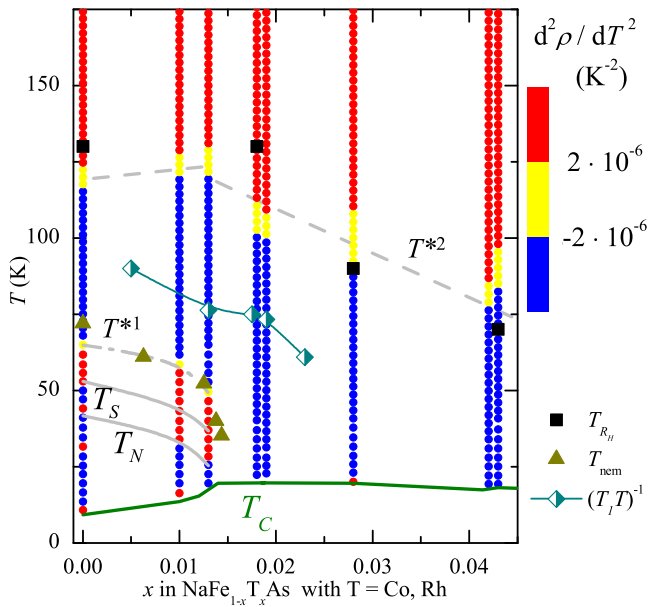}
\caption{(Color online) Color-coded phase diagram of NaFe$_{1-x}$T$_x$ with T~=~Co and Rh. The curvature of the resistivity of the \rnf single crystals is shown. Red color marks a positive, yellow nearly no curvature and blue color a negative curvature of the resistivity data sets. The grey dotted lines are guides to the eye to mark the transition regions $T^{*1}$ and $T^{*2}$. The grey solid and green lines mark the phase transition temperatures from Ref.~\onlinecite{Steckel2015}. The dark yellow triangles mark the nematic transition temperature found by Ref.~\onlinecite{Deng2015} by analyzing the resistivity anisotropy of detwinned \cnfy single crystals with the nominal Co-doping level $y$ rescaled to match optimal doping of our samples with $x=0.018$. The half-open turquoise dots mark the $1/T_1T=0.22$~(sK)$^{-1}$ data points from NMR measurements\cite{Ma2014} on \cnf single crystals (see Discussion). The black squares mark the deviation temperature $T_{R_H}$ at which the Hall coefficient deviates from the high temperature phenomenological fit.}
\label{fig:2}
\end{figure}

The inflection point in the electrical resistivity curves is known as indicator for changes of the electronic structure. \cite{Ando2004, Hess2009} Thus, to investigate the temperature regime of changes in the electronic structure in more detail we plot the curvature of the resistivity data given by the second derivative in a color-coded scheme in Fig.~\ref{fig:2}. In particular, we identify the inflection points by zero curvature.

The phase diagram yields two inflection point regimes. A first one $T^{*1}$ in the underdoped regime at $\sim 20$~K higher than \ts. The structural transition seems to follow $T^{*1}$ and both vanish upon doping towards the optimal doping level. The second inflection point regime $T^{*2}$ is remarkably high in temperature. At first it increases slightly from 125~K in the undoped NaFeAs up to 125~K in the optimally doped ($x=0.013$) crystal, then $T^{*2}$ decreases down to 75~K with further increasing $x$ up to the highest doping levels, i.e. $T^{*2}(x)$ changes slope at about optimal doping. Thus, in contrast to $T^{*1}$, the second inflection point regime is traceable through the whole accessible phase diagram.

\subsection{Hall effect}

\begin{figure}[!t]
\includegraphics[clip,width=1\columnwidth]{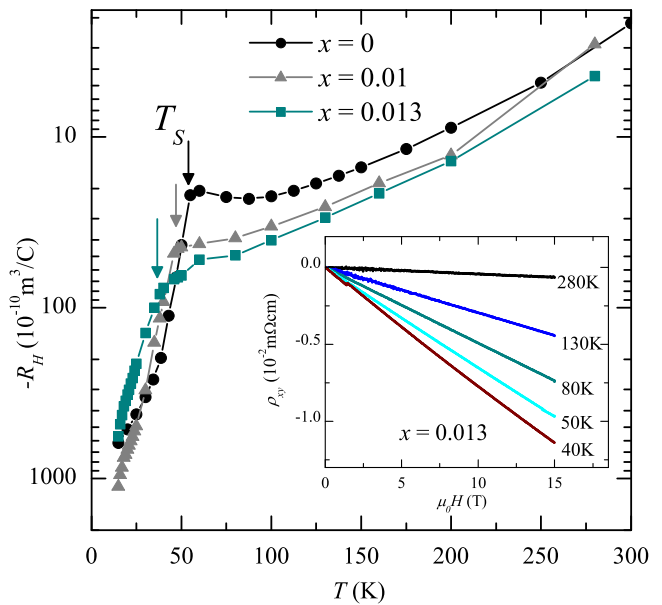}
\caption{(Color online) Hall coefficient $R_H$ from the underdoped \rnf crystals at 8~T. The arrow marks the structural transition temperature $T_S$ from Ref.~\onlinecite{Steckel2015}. The inset shows the diagonal resistivity $\rho_{xy}$ in dependence of the absolute magnetic field of the $x=0.013$ underdoped sample.}
\label{fig:3}
\end{figure}

\begin{figure}
\includegraphics[clip,width=1\columnwidth]{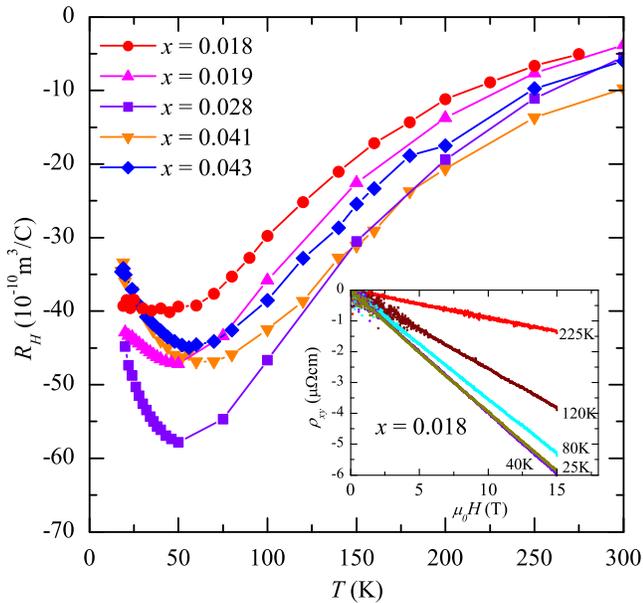}
\caption{(Color online) Hall coefficient $R_H$ from the optimal and overdoped \rnf crystals. The inset shows the diagonal resistivity $\rho_{xy}$ in dependence of the absolute magnetic field of the $x=0.018$ optimally doped sample.}
\label{fig:4}
\end{figure}

The Hall coefficient $R_H$ (see Fig.~\ref{fig:3} and \ref{fig:4}) is calculated from the slope of $\rho_{xy}(|B|)$. Please note that the undoped NaFeAs has a nonlinear $\rho_{xy}$ (not shown) in agreement with an earlier report \cite{Deng2015a}. $R_H$ of NaFeAs is negative in the complete measured temperature range and the absolute value $|R_H|$ increases weakly with decreasing temperature. At $T\geq T_S$ the temperature dependence of $R_H$ is relatively small while at the structural transition temperature $T_S$ the Hall coefficient has a kink and rises strongly to high negative values without any further anomaly down to the lowest accessible temperatures. Especially, at the magnetic ordering temperature no anomaly is revealed. This shape of $R_H$ is similar for the underdoped crystals $x\leq 0.013$ (Fig.~\ref{fig:3}) except for a shifted and smeared structural phase transition anomaly in analogy to the magnetic susceptibility $\chi$. For Rh contents higher than $x=0.013$ the nonlinearity of $\rho_{xy}(|B|)$ vanishes and no kink appears which provides more evidence that the optimally doped \rnf sample with $x=0.018$ has no structural transition. \cite{Steckel2015} The origin of the kink can be understood with the help of temperature-dependent ARPES measurements on NaFeAs.\cite{He2010} These data have shown that a big part of the band structure starts to shift at \ts prior to the electronic reconstruction due to the magnetic ordering. Such a reconstruction at the Fermi surface involves naturally a strong change of the charge carrier density and, thus, a direct signal in the Hall coefficient.

The $R_H$ of the overdoped samples, i.e. $x\geq 0.018$, plotted in Fig.~\ref{fig:4} has nearly the same weak temperature dependence as the underdoped crystals down to $\sim 50$~K where $|R_H|$ has a maximum and decreases for lower temperatures. This temperature dependence is highly comparable to that measured on Co-doped NaFeAs. \cite{Chen2009,Wang2013a,Wang2013} 

An interesting quantity to get more insight to the temperature dependence of the multiband Hall coefficient is the cotangent of the Hall angle which is defined as $\cot \theta_H=\rho_{xx}/\rho_{xy}$. In the Drude one-band picture, where $R_H=1/nq$ and $\rho_{xx}=m/nq^2\tau$, it follows:
\begin{equation}
\cot \theta_H = \frac{\rho}{R_HB}=\frac{m}{q\tau B}.
\label{eq:cot}
\end{equation} 
In this quantity the influence of the charge carrier density $n$ cancels, and the cotangent of the Hall angle is direct proportional to the effective scattering rate. In the case of more than one contributing band to the charge carrier transport the dependencies are not easily revealed. Nevertheless, in high-temperature superconductors typically a phenomenological function of the form
\begin{equation}
\cot \theta_H = a + b T^\beta
\label{eq:beta}
\end{equation}
fits the temperature dependence of the cotangent of the Hall angle very well.\cite{Chien1991, Anderson1991, Yan2013, Arushanov2011, Wang2013a} In the multiband Fe-based superconductors, in particular doped BaFe$_2$As$_2$ \cite{Yan2013, Arushanov2011} and Co-doped NaFeAs \cite{Wang2013a} $\beta$-values between 4 and 2 have been reported. Motivated by these findings we address now the analysis of the Hall effect data on our \rnf single crystals.

\begin{figure}
\includegraphics[clip,width=1\columnwidth]{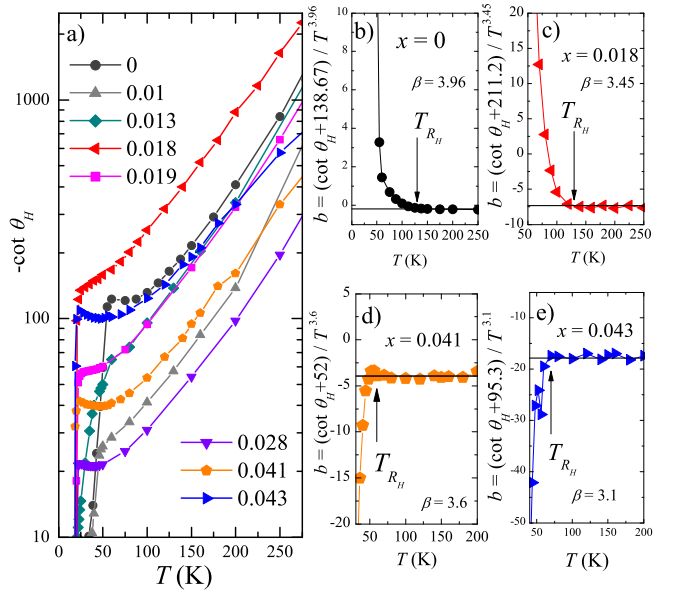}
\caption{(Color online) The negative cotangent of the Hall angle $\theta_H$ for all \rnf single crystals in a semilogarithmic plot (a). On the right side (b)-(e), for selected compositions, the deviation from the high temperature fit by Eq.~(\ref{eq:beta}), plotted as $b=(\cot \theta_H-a)/T^\beta$. The fit range was chosen as $T\geq 100$~K. $T_{R_H}$ marks the temperature, where the data points deviate from the high temperature fit.}
\label{fig:5}
\end{figure}

The cotangent of the Hall angle of the \rnf single crystals is displayed in Fig.~\ref{fig:5}(a). The high temperature behavior ($T>125$~K) is well described by Eq.~(\ref{eq:beta}). However, below a certain temperature the curves shown in Fig.~\ref{fig:5}(b)-(e) deviate from this power law. To illustrate this behavior, we subtracted the high temperature fit from Eq.~(\ref{eq:beta}) from the data points to clearly define the deviation temperature $T_{R_H}$ (cf. Fig.~\ref{fig:5}). Additionally, the values for $\beta$ are given in the plots and they vary between 4 and 3, while $\beta$ consistently with previous data sets,\cite{Yan2013, Arushanov2011, Wang2013a} decreases with increasing doping level. 

Nevertheless, such a deviation from a temperature law which is valid for a larger temperature regime, points towards an unusual change in the physical properties of the charge carriers, i.e. effective mass or scattering rate following Eq.~(\ref{eq:cot}). Indeed, also the charge carrier density $n$ can be responsible for the changes, because Eq.~(\ref{eq:cot}) is strictly valid only in a one-band metal.

Remarkably, the Hall coefficient deviation temperatures $T_{R_H}$, additionally plotted in Fig.~\ref{fig:2}, reflect as well the aforementioned trend of the second inflection point region $T^{*2}$. Thus, we have a second, independent determination of $T^{*2}$. Additionally, the comparison with Co-doped NaFeAs shows that both materials have a similar transition region as well as similar phase transition temperatures upon formal electron doping.

\section{Discussion}

After having established the experimental finding of two transition regions $T^{*1}$ and $T^{*2}$ in the electronic phase diagram of electron-doped NaFeAs above the known phase transition temperatures the natural question of the origin of these inflection points has to be answered. Therefore, we discuss our results in the light of other experimental results for the tetragonal phase in this material.
 
In a resistivity anisotropy study on detwinned \cnf in Ref.~\onlinecite{Deng2015} the onset of the anisotropy is defined as the kink appearing in $\rho_a-\rho_b$ slightly above \ts. The onset temperatures $T_\mathrm{nem}$ of the resistivity anisotropy of \cnf nearly coincides with $T^{*1}$ in unstrained \rnf single crystals (cf. Fig.~\ref{fig:2}). Thus, it is possible to track the onset temperature $T_\mathrm{nem}$ of a strong anisotropy between $\rho_a$ and $\rho_b$ by carefully studying the stress-free average resistivity curvature. Note that, we found small deviations especially in the undoped crystal where $T_\mathrm{nem}$ is located at a slightly higher temperature than $T^{*1}$, which might either have its origin in the uncertainty of $\delta$ in Na$_{1-\delta}$FeAs or in the smearing of the structural phase transition by uniaxially stressing the crystal. All the shown data, i.e. $T_\mathrm{nem}$ and $T^{*1}$ have in common, that they are tightly connected to the structural transition and are suppressed upon doping equally to \ts.

We ascribed the second inflection point $T^{*2}$ at much higher temperatures to a change of either the charge carrier density, the effective mass, the scattering rate or a combination of those. From the $\cot \theta_H$ analysis we know that this transition region cannot be assigned to the multiband nature of the Fe-based superconductors. Besides, no other phase transition at such high temperatures in these materials are known. Another correlation could be the influence of fluctuations on the transport which could be fluctuations of the spins, the orbitals or the structure. We compare our results with spin fluctuations in the tetragonal state and therefore consult NMR data of Co-doped NaFeAs, in particular we use the quantity $1/T_1T$.\cite{Ma2014} This quantity is proportional to the imaginary part of the dynamical spin susceptibility and thus measures spin fluctuations in the whole Brillouing zone. $1/T_1T$ rises significantly far above the structural transition temperature and thus reveals a slow-down of the spin fluctuations. By definition, this includes $q=0$ nematic fluctuations and indeed a scaling of $1/T_1T$ \cite{Ning2010} with the elastic moduli, \cite{Bohmer2014} showing the softness of the crystal lattice above \ts, has been reported.\cite{Fernandes2013} For a proper comparison we choose a certain $1/T_1T$ value (see Fig.~\ref{fig:2}) and mark the temperatures at which the $1/T_1T$ of a particular Co-doped NaFeAs crystal crosses this value. Interestingly, the doping dependence in the electronic phase diagram of $T^{*2}$ and of the $1/T_1T$ values is similar. In particular, $T^{*2}$ and $1/T_1T$ as a function of doping have the same slope for doping levels above optimal doping. We therefore ascribe the inflection point to a sensitivity of the resistivity to the onset of nematic fluctuations. It still remains to be clarified which quantities, i.e. charge carrier density, effective mass and scattering rate, are dominantly influenced. Conflicting results about the importance of the impurity density and their anisotropy have been reported. While transport measurements in magnetic field\cite{Deng2015a} and after annealing\cite{Ishida2013} suggest a dominant role of the observed anisotropic impurity states\cite{Allan2013} another strain dependent resistivity anisotropy experiment points towards a negligible influence of the impurity density above \ts.\cite{Kuo2014}

We point out that the electrical transport coefficients seem to be quite more sensitive to nematic fluctuations in the tetragonal phase of the iron-based superconductors than other probes such as NMR and STM\cite{Rosenthal2014} experiments, which found anisotropies up to 90~K. In view of this strong sensitivity it is remarkable that we can resolve the aforementioned slope change of $T^{*2}(x)$ at optimal doping. Such an anomaly in the doping dependence at optimal doping has been reported before in resistivity anisotropy measurements.\cite{Kuo2011, Chu2010} The corroboration of these findings by our results calls for a further investigation of the fluctuation regime to disentangle whether the nematic fluctuation channels or the coupling constants\cite{Blomberg2013} possess a hidden doping dependence including a clarification of the role of the impurity density in the nematic fluctuation regime.

\section{Conclusion}
We performed a combined study of resistivity and Hall effect measurements on \rnf single crystals. In total we found the typical anomalies of the phase transitions and additional deviations from the expected high temperature behavior at temperatures far above the phase transitions. We applied the method of the resistivity curvature analysis and found two regions of inflection points. The first inflection point $T^{*1}$ at temperatures slightly above \ts points towards the onset of resistivity anisotropy. Comparisons to NMR data suggest that $T^{*2}$ from the resistivity as well as the Hall coefficient indicate the onset of fluctuations connected to the nematic phase in the tetragonal state. This method is thus capable of revealing not only the broad fluctuation regime at high temperatures in the complete phase diagram but also the real onset of resistivity anisotropy induced by the nematic rotational symmetry breaking which is connected with the structural phase transition.

\section*{ACKNOWLEDGMENTS}
This work has been supported by the Deutsche Forschungsgemeinschaft through the Priority Program SPP1458 (Grants No. BU887/15-1 and HE3439/11), and through the Emmy Noether Program in project WU595/3-3. We thank the BMBF for support in the frame of the ERA.Net RUS project (project 01DJ12096, FeSuCo).


\begin{thebibliography}{45}%
\makeatletter
\providecommand \@ifxundefined [1]{%
 \@ifx{#1\undefined}
}%
\providecommand \@ifnum [1]{%
 \ifnum #1\expandafter \@firstoftwo
 \else \expandafter \@secondoftwo
 \fi
}%
\providecommand \@ifx [1]{%
 \ifx #1\expandafter \@firstoftwo
 \else \expandafter \@secondoftwo
 \fi
}%
\providecommand \natexlab [1]{#1}%
\providecommand \enquote  [1]{``#1''}%
\providecommand \bibnamefont  [1]{#1}%
\providecommand \bibfnamefont [1]{#1}%
\providecommand \citenamefont [1]{#1}%
\providecommand \href@noop [0]{\@secondoftwo}%
\providecommand \href [0]{\begingroup \@sanitize@url \@href}%
\providecommand \@href[1]{\@@startlink{#1}\@@href}%
\providecommand \@@href[1]{\endgroup#1\@@endlink}%
\providecommand \@sanitize@url [0]{\catcode `\\12\catcode `\$12\catcode
  `\&12\catcode `\#12\catcode `\^12\catcode `\_12\catcode `\%12\relax}%
\providecommand \@@startlink[1]{}%
\providecommand \@@endlink[0]{}%
\providecommand \url  [0]{\begingroup\@sanitize@url \@url }%
\providecommand \@url [1]{\endgroup\@href {#1}{\urlprefix }}%
\providecommand \urlprefix  [0]{URL }%
\providecommand \Eprint [0]{\href }%
\providecommand \doibase [0]{http://dx.doi.org/}%
\providecommand \selectlanguage [0]{\@gobble}%
\providecommand \bibinfo  [0]{\@secondoftwo}%
\providecommand \bibfield  [0]{\@secondoftwo}%
\providecommand \translation [1]{[#1]}%
\providecommand \BibitemOpen [0]{}%
\providecommand \bibitemStop [0]{}%
\providecommand \bibitemNoStop [0]{.\EOS\space}%
\providecommand \EOS [0]{\spacefactor3000\relax}%
\providecommand \BibitemShut  [1]{\csname bibitem#1\endcsname}%
\let\auto@bib@innerbib\@empty
\bibitem [{\citenamefont {Steckel}\ \emph {et~al.}(2015)\citenamefont
  {Steckel}, \citenamefont {Roslova}, \citenamefont {Beck}, \citenamefont
  {Morozov}, \citenamefont {Aswartham}, \citenamefont {Evtushinsky},
  \citenamefont {Blum}, \citenamefont {Abdel-Hafiez}, \citenamefont {Bombor},
  \citenamefont {Maletz}, \citenamefont {Borisenko}, \citenamefont {Shevelkov},
  \citenamefont {Wolter}, \citenamefont {Hess}, \citenamefont {Wurmehl},\ and\
  \citenamefont {B\"uchner}}]{Steckel2015}%
  \BibitemOpen
  \bibfield  {author} {\bibinfo {author} {\bibfnamefont {F.}~\bibnamefont
  {Steckel}}, \bibinfo {author} {\bibfnamefont {M.}~\bibnamefont {Roslova}},
  \bibinfo {author} {\bibfnamefont {R.}~\bibnamefont {Beck}}, \bibinfo {author}
  {\bibfnamefont {I.}~\bibnamefont {Morozov}}, \bibinfo {author} {\bibfnamefont
  {S.}~\bibnamefont {Aswartham}}, \bibinfo {author} {\bibfnamefont
  {D.}~\bibnamefont {Evtushinsky}}, \bibinfo {author} {\bibfnamefont
  {C.~G.~F.}\ \bibnamefont {Blum}}, \bibinfo {author} {\bibfnamefont
  {M.}~\bibnamefont {Abdel-Hafiez}}, \bibinfo {author} {\bibfnamefont
  {D.}~\bibnamefont {Bombor}}, \bibinfo {author} {\bibfnamefont
  {J.}~\bibnamefont {Maletz}}, \bibinfo {author} {\bibfnamefont
  {S.}~\bibnamefont {Borisenko}}, \bibinfo {author} {\bibfnamefont {A.~V.}\
  \bibnamefont {Shevelkov}}, \bibinfo {author} {\bibfnamefont {A.~U.~B.}\
  \bibnamefont {Wolter}}, \bibinfo {author} {\bibfnamefont {C.}~\bibnamefont
  {Hess}}, \bibinfo {author} {\bibfnamefont {S.}~\bibnamefont {Wurmehl}}, \
  and\ \bibinfo {author} {\bibfnamefont {B.}~\bibnamefont {B\"uchner}},\ }\href
  {\doibase 10.1103/PhysRevB.91.184516} {\bibfield  {journal} {\bibinfo
  {journal} {Phys. Rev. B}\ }\textbf {\bibinfo {volume} {91}},\ \bibinfo
  {pages} {184516} (\bibinfo {year} {2015})}\BibitemShut {NoStop}%
\bibitem [{\citenamefont {Ni}\ \emph {et~al.}(2009)\citenamefont {Ni},
  \citenamefont {Thaler}, \citenamefont {Kracher}, \citenamefont {Yan},
  \citenamefont {Bud'ko},\ and\ \citenamefont {Canfield}}]{Ni2009}%
  \BibitemOpen
  \bibfield  {author} {\bibinfo {author} {\bibfnamefont {N.}~\bibnamefont
  {Ni}}, \bibinfo {author} {\bibfnamefont {A.}~\bibnamefont {Thaler}}, \bibinfo
  {author} {\bibfnamefont {A.}~\bibnamefont {Kracher}}, \bibinfo {author}
  {\bibfnamefont {J.~Q.}\ \bibnamefont {Yan}}, \bibinfo {author} {\bibfnamefont
  {S.~L.}\ \bibnamefont {Bud'ko}}, \ and\ \bibinfo {author} {\bibfnamefont
  {P.~C.}\ \bibnamefont {Canfield}},\ }\href {\doibase
  10.1103/PhysRevB.80.024511} {\bibfield  {journal} {\bibinfo  {journal} {Phys.
  Rev. B}\ }\textbf {\bibinfo {volume} {80}},\ \bibinfo {pages} {024511}
  (\bibinfo {year} {2009})}\BibitemShut {NoStop}%
\bibitem [{\citenamefont {Fernandes}\ \emph {et~al.}(2014)\citenamefont
  {Fernandes}, \citenamefont {Chubukov},\ and\ \citenamefont
  {Schmalian}}]{Fernandes2014}%
  \BibitemOpen
  \bibfield  {author} {\bibinfo {author} {\bibfnamefont {R.~M.}\ \bibnamefont
  {Fernandes}}, \bibinfo {author} {\bibfnamefont {A.~V.}\ \bibnamefont
  {Chubukov}}, \ and\ \bibinfo {author} {\bibfnamefont {J.}~\bibnamefont
  {Schmalian}},\ }\href {\doibase 10.1038/nphys2877} {\bibfield  {journal}
  {\bibinfo  {journal} {Nat. Phys.}\ }\textbf {\bibinfo {volume} {10}},\
  \bibinfo {pages} {97} (\bibinfo {year} {2014})}\BibitemShut {NoStop}%
\bibitem [{\citenamefont {B\"ohmer}\ and\ \citenamefont
  {Meingast}(2015)}]{Boehmer2015}%
  \BibitemOpen
  \bibfield  {author} {\bibinfo {author} {\bibfnamefont {A.~E.}\ \bibnamefont
  {B\"ohmer}}\ and\ \bibinfo {author} {\bibfnamefont {C.}~\bibnamefont
  {Meingast}},\ }\href {\doibase 10.1016/j.crhy.2015.07.001} {\bibfield
  {journal} {\bibinfo  {journal} {C. R. Physique}\ } (\bibinfo {year} {2015}),\
  10.1016/j.crhy.2015.07.001}\BibitemShut {NoStop}%
\bibitem [{\citenamefont {Kivelson}\ \emph {et~al.}(1998)\citenamefont
  {Kivelson}, \citenamefont {Fradkin},\ and\ \citenamefont
  {Emery}}]{Kivelson1998}%
  \BibitemOpen
  \bibfield  {author} {\bibinfo {author} {\bibfnamefont {S.~A.}\ \bibnamefont
  {Kivelson}}, \bibinfo {author} {\bibfnamefont {E.}~\bibnamefont {Fradkin}}, \
  and\ \bibinfo {author} {\bibfnamefont {V.~J.}\ \bibnamefont {Emery}},\ }\href
  {\doibase 10.1038/31177} {\bibfield  {journal} {\bibinfo  {journal} {Nature}\
  }\textbf {\bibinfo {volume} {393}},\ \bibinfo {pages} {550} (\bibinfo {year}
  {1998})}\BibitemShut {NoStop}%
\bibitem [{\citenamefont {Rosenthal}\ \emph {et~al.}(2014)\citenamefont
  {Rosenthal}, \citenamefont {Andrade}, \citenamefont {Arguello}, \citenamefont
  {Fernandes}, \citenamefont {Xing}, \citenamefont {Wang}, \citenamefont {Jin},
  \citenamefont {Millis},\ and\ \citenamefont {Pasupathy}}]{Rosenthal2014}%
  \BibitemOpen
  \bibfield  {author} {\bibinfo {author} {\bibfnamefont {E.~P.}\ \bibnamefont
  {Rosenthal}}, \bibinfo {author} {\bibfnamefont {E.~F.}\ \bibnamefont
  {Andrade}}, \bibinfo {author} {\bibfnamefont {C.~J.}\ \bibnamefont
  {Arguello}}, \bibinfo {author} {\bibfnamefont {R.~M.}\ \bibnamefont
  {Fernandes}}, \bibinfo {author} {\bibfnamefont {L.~Y.}\ \bibnamefont {Xing}},
  \bibinfo {author} {\bibfnamefont {X.~C.}\ \bibnamefont {Wang}}, \bibinfo
  {author} {\bibfnamefont {C.~Q.}\ \bibnamefont {Jin}}, \bibinfo {author}
  {\bibfnamefont {A.~J.}\ \bibnamefont {Millis}}, \ and\ \bibinfo {author}
  {\bibfnamefont {A.~N.}\ \bibnamefont {Pasupathy}},\ }\href {\doibase
  10.1038/nphys2870} {\bibfield  {journal} {\bibinfo  {journal} {Nat. Phys.}\
  }\textbf {\bibinfo {volume} {10}},\ \bibinfo {pages} {225} (\bibinfo {year}
  {2014})}\BibitemShut {NoStop}%
\bibitem [{\citenamefont {Chuang}\ \emph {et~al.}(2010)\citenamefont {Chuang},
  \citenamefont {Allan}, \citenamefont {Lee}, \citenamefont {Xie},
  \citenamefont {Ni}, \citenamefont {Bud?ko}, \citenamefont {Boebinger},
  \citenamefont {Canfield},\ and\ \citenamefont {Davis}}]{Chuang2010}%
  \BibitemOpen
  \bibfield  {author} {\bibinfo {author} {\bibfnamefont {T.-M.}\ \bibnamefont
  {Chuang}}, \bibinfo {author} {\bibfnamefont {M.~P.}\ \bibnamefont {Allan}},
  \bibinfo {author} {\bibfnamefont {J.}~\bibnamefont {Lee}}, \bibinfo {author}
  {\bibfnamefont {Y.}~\bibnamefont {Xie}}, \bibinfo {author} {\bibfnamefont
  {N.}~\bibnamefont {Ni}}, \bibinfo {author} {\bibfnamefont {S.~L.}\
  \bibnamefont {Bud?ko}}, \bibinfo {author} {\bibfnamefont {G.~S.}\
  \bibnamefont {Boebinger}}, \bibinfo {author} {\bibfnamefont {P.~C.}\
  \bibnamefont {Canfield}}, \ and\ \bibinfo {author} {\bibfnamefont {J.~C.}\
  \bibnamefont {Davis}},\ }\href {\doibase 10.1126/science.1181083} {\bibfield
  {journal} {\bibinfo  {journal} {Science}\ }\textbf {\bibinfo {volume}
  {327}},\ \bibinfo {pages} {181} (\bibinfo {year} {2010})}\BibitemShut
  {NoStop}%
\bibitem [{\citenamefont {Blomberg}\ \emph {et~al.}(2013)\citenamefont
  {Blomberg}, \citenamefont {Tanatar}, \citenamefont {Fernandes}, \citenamefont
  {Mazin}, \citenamefont {Shen}, \citenamefont {Wen}, \citenamefont {Johannes},
  \citenamefont {Schmalian},\ and\ \citenamefont {Prozorov}}]{Blomberg2013}%
  \BibitemOpen
  \bibfield  {author} {\bibinfo {author} {\bibfnamefont {E.~C.}\ \bibnamefont
  {Blomberg}}, \bibinfo {author} {\bibfnamefont {M.~A.}\ \bibnamefont
  {Tanatar}}, \bibinfo {author} {\bibfnamefont {R.~M.}\ \bibnamefont
  {Fernandes}}, \bibinfo {author} {\bibfnamefont {I.~I.}\ \bibnamefont
  {Mazin}}, \bibinfo {author} {\bibfnamefont {B.}~\bibnamefont {Shen}},
  \bibinfo {author} {\bibfnamefont {H.-H.}\ \bibnamefont {Wen}}, \bibinfo
  {author} {\bibfnamefont {M.~D.}\ \bibnamefont {Johannes}}, \bibinfo {author}
  {\bibfnamefont {J.}~\bibnamefont {Schmalian}}, \ and\ \bibinfo {author}
  {\bibfnamefont {R.}~\bibnamefont {Prozorov}},\ }\href {\doibase
  10.1038/ncomms2933} {\bibfield  {journal} {\bibinfo  {journal} {Nat.
  Commun.}\ }\textbf {\bibinfo {volume} {4}},\ \bibinfo {pages} {1914}
  (\bibinfo {year} {2013})}\BibitemShut {NoStop}%
\bibitem [{\citenamefont {Deng}\ \emph
  {et~al.}(2015{\natexlab{a}})\citenamefont {Deng}, \citenamefont {Liu},
  \citenamefont {Xing}, \citenamefont {Yang},\ and\ \citenamefont
  {Wen}}]{Deng2015}%
  \BibitemOpen
  \bibfield  {author} {\bibinfo {author} {\bibfnamefont {Q.}~\bibnamefont
  {Deng}}, \bibinfo {author} {\bibfnamefont {J.}~\bibnamefont {Liu}}, \bibinfo
  {author} {\bibfnamefont {J.}~\bibnamefont {Xing}}, \bibinfo {author}
  {\bibfnamefont {H.}~\bibnamefont {Yang}}, \ and\ \bibinfo {author}
  {\bibfnamefont {H.-H.}\ \bibnamefont {Wen}},\ }\href {\doibase
  10.1103/PhysRevB.91.020508} {\bibfield  {journal} {\bibinfo  {journal} {Phys.
  Rev. B}\ }\textbf {\bibinfo {volume} {91}},\ \bibinfo {pages} {020508}
  (\bibinfo {year} {2015}{\natexlab{a}})}\BibitemShut {NoStop}%
\bibitem [{\citenamefont {Chu}\ \emph {et~al.}(2010)\citenamefont {Chu},
  \citenamefont {Analytis}, \citenamefont {Greve}, \citenamefont {McMahon},
  \citenamefont {Islam}, \citenamefont {Yamamoto},\ and\ \citenamefont
  {Fisher}}]{Chu2010}%
  \BibitemOpen
  \bibfield  {author} {\bibinfo {author} {\bibfnamefont {J.-H.}\ \bibnamefont
  {Chu}}, \bibinfo {author} {\bibfnamefont {J.~G.}\ \bibnamefont {Analytis}},
  \bibinfo {author} {\bibfnamefont {K.~D.}\ \bibnamefont {Greve}}, \bibinfo
  {author} {\bibfnamefont {P.~L.}\ \bibnamefont {McMahon}}, \bibinfo {author}
  {\bibfnamefont {Z.}~\bibnamefont {Islam}}, \bibinfo {author} {\bibfnamefont
  {Y.}~\bibnamefont {Yamamoto}}, \ and\ \bibinfo {author} {\bibfnamefont
  {I.~R.}\ \bibnamefont {Fisher}},\ }\href {\doibase 10.1126/science.1190482}
  {\bibfield  {journal} {\bibinfo  {journal} {Science}\ }\textbf {\bibinfo
  {volume} {329}},\ \bibinfo {pages} {824} (\bibinfo {year}
  {2010})}\BibitemShut {NoStop}%
\bibitem [{\citenamefont {Fisher}\ \emph {et~al.}(2011)\citenamefont {Fisher},
  \citenamefont {Degiorgi},\ and\ \citenamefont {Shen}}]{Fisher2011}%
  \BibitemOpen
  \bibfield  {author} {\bibinfo {author} {\bibfnamefont {I.~R.}\ \bibnamefont
  {Fisher}}, \bibinfo {author} {\bibfnamefont {L.}~\bibnamefont {Degiorgi}}, \
  and\ \bibinfo {author} {\bibfnamefont {Z.~X.}\ \bibnamefont {Shen}},\ }\href
  {\doibase 10.1088/0034-4885/74/12/124506} {\bibfield  {journal} {\bibinfo
  {journal} {Rep. Prog. Phys}\ }\textbf {\bibinfo {volume} {74}},\ \bibinfo
  {pages} {124506} (\bibinfo {year} {2011})}\BibitemShut {NoStop}%
\bibitem [{\citenamefont {Jiang}\ \emph {et~al.}(2013)\citenamefont {Jiang},
  \citenamefont {Jeevan}, \citenamefont {Dong},\ and\ \citenamefont
  {Gegenwart}}]{Jiang2013}%
  \BibitemOpen
  \bibfield  {author} {\bibinfo {author} {\bibfnamefont {S.}~\bibnamefont
  {Jiang}}, \bibinfo {author} {\bibfnamefont {H.~S.}\ \bibnamefont {Jeevan}},
  \bibinfo {author} {\bibfnamefont {J.}~\bibnamefont {Dong}}, \ and\ \bibinfo
  {author} {\bibfnamefont {P.}~\bibnamefont {Gegenwart}},\ }\href {\doibase
  10.1103/PhysRevLett.110.067001} {\bibfield  {journal} {\bibinfo  {journal}
  {Phys. Rev. Lett.}\ }\textbf {\bibinfo {volume} {110}},\ \bibinfo {pages}
  {067001} (\bibinfo {year} {2013})}\BibitemShut {NoStop}%
\bibitem [{\citenamefont {Tanatar}\ \emph {et~al.}(2010)\citenamefont
  {Tanatar}, \citenamefont {Blomberg}, \citenamefont {Kreyssig}, \citenamefont
  {Kim}, \citenamefont {Ni}, \citenamefont {Thaler}, \citenamefont {Bud'ko},
  \citenamefont {Canfield}, \citenamefont {Goldman}, \citenamefont {Mazin},\
  and\ \citenamefont {Prozorov}}]{Tanatar2010}%
  \BibitemOpen
  \bibfield  {author} {\bibinfo {author} {\bibfnamefont {M.~A.}\ \bibnamefont
  {Tanatar}}, \bibinfo {author} {\bibfnamefont {E.~C.}\ \bibnamefont
  {Blomberg}}, \bibinfo {author} {\bibfnamefont {A.}~\bibnamefont {Kreyssig}},
  \bibinfo {author} {\bibfnamefont {M.~G.}\ \bibnamefont {Kim}}, \bibinfo
  {author} {\bibfnamefont {N.}~\bibnamefont {Ni}}, \bibinfo {author}
  {\bibfnamefont {A.}~\bibnamefont {Thaler}}, \bibinfo {author} {\bibfnamefont
  {S.~L.}\ \bibnamefont {Bud'ko}}, \bibinfo {author} {\bibfnamefont {P.~C.}\
  \bibnamefont {Canfield}}, \bibinfo {author} {\bibfnamefont {A.~I.}\
  \bibnamefont {Goldman}}, \bibinfo {author} {\bibfnamefont {I.~I.}\
  \bibnamefont {Mazin}}, \ and\ \bibinfo {author} {\bibfnamefont
  {R.}~\bibnamefont {Prozorov}},\ }\href {\doibase 10.1103/PhysRevB.81.184508}
  {\bibfield  {journal} {\bibinfo  {journal} {Phys. Rev. B}\ }\textbf {\bibinfo
  {volume} {81}},\ \bibinfo {pages} {184508} (\bibinfo {year}
  {2010})}\BibitemShut {NoStop}%
\bibitem [{\citenamefont {Ying}\ \emph {et~al.}(2011)\citenamefont {Ying},
  \citenamefont {Wang}, \citenamefont {Wu}, \citenamefont {Xiang},
  \citenamefont {Liu}, \citenamefont {Yan}, \citenamefont {Wang}, \citenamefont
  {Zhang}, \citenamefont {Ye}, \citenamefont {Cheng}, \citenamefont {Hu},\ and\
  \citenamefont {Chen}}]{Ying2011}%
  \BibitemOpen
  \bibfield  {author} {\bibinfo {author} {\bibfnamefont {J.~J.}\ \bibnamefont
  {Ying}}, \bibinfo {author} {\bibfnamefont {X.~F.}\ \bibnamefont {Wang}},
  \bibinfo {author} {\bibfnamefont {T.}~\bibnamefont {Wu}}, \bibinfo {author}
  {\bibfnamefont {Z.~J.}\ \bibnamefont {Xiang}}, \bibinfo {author}
  {\bibfnamefont {R.~H.}\ \bibnamefont {Liu}}, \bibinfo {author} {\bibfnamefont
  {Y.~J.}\ \bibnamefont {Yan}}, \bibinfo {author} {\bibfnamefont {A.~F.}\
  \bibnamefont {Wang}}, \bibinfo {author} {\bibfnamefont {M.}~\bibnamefont
  {Zhang}}, \bibinfo {author} {\bibfnamefont {G.~J.}\ \bibnamefont {Ye}},
  \bibinfo {author} {\bibfnamefont {P.}~\bibnamefont {Cheng}}, \bibinfo
  {author} {\bibfnamefont {J.~P.}\ \bibnamefont {Hu}}, \ and\ \bibinfo {author}
  {\bibfnamefont {X.~H.}\ \bibnamefont {Chen}},\ }\href {\doibase
  10.1103/PhysRevLett.107.067001} {\bibfield  {journal} {\bibinfo  {journal}
  {Phys. Rev. Lett.}\ }\textbf {\bibinfo {volume} {107}},\ \bibinfo {pages}
  {067001} (\bibinfo {year} {2011})}\BibitemShut {NoStop}%
\bibitem [{\citenamefont {Ma}\ \emph {et~al.}(2014)\citenamefont {Ma},
  \citenamefont {Dai}, \citenamefont {Wang}, \citenamefont {Lu}, \citenamefont
  {Song}, \citenamefont {Zhang}, \citenamefont {Tan}, \citenamefont {Dai},
  \citenamefont {Hu}, \citenamefont {Li}, \citenamefont {Normand},\ and\
  \citenamefont {Yu}}]{Ma2014}%
  \BibitemOpen
  \bibfield  {author} {\bibinfo {author} {\bibfnamefont {L.}~\bibnamefont
  {Ma}}, \bibinfo {author} {\bibfnamefont {J.}~\bibnamefont {Dai}}, \bibinfo
  {author} {\bibfnamefont {P.~S.}\ \bibnamefont {Wang}}, \bibinfo {author}
  {\bibfnamefont {X.~R.}\ \bibnamefont {Lu}}, \bibinfo {author} {\bibfnamefont
  {Y.}~\bibnamefont {Song}}, \bibinfo {author} {\bibfnamefont {C.}~\bibnamefont
  {Zhang}}, \bibinfo {author} {\bibfnamefont {G.~T.}\ \bibnamefont {Tan}},
  \bibinfo {author} {\bibfnamefont {P.}~\bibnamefont {Dai}}, \bibinfo {author}
  {\bibfnamefont {D.}~\bibnamefont {Hu}}, \bibinfo {author} {\bibfnamefont
  {S.~L.}\ \bibnamefont {Li}}, \bibinfo {author} {\bibfnamefont
  {B.}~\bibnamefont {Normand}}, \ and\ \bibinfo {author} {\bibfnamefont
  {W.}~\bibnamefont {Yu}},\ }\href {\doibase 10.1103/PhysRevB.90.144502}
  {\bibfield  {journal} {\bibinfo  {journal} {Phys. Rev. B}\ }\textbf {\bibinfo
  {volume} {90}},\ \bibinfo {pages} {144502} (\bibinfo {year}
  {2014})}\BibitemShut {NoStop}%
\bibitem [{\citenamefont {Ren}\ \emph {et~al.}(2015)\citenamefont {Ren},
  \citenamefont {Duan}, \citenamefont {Hu}, \citenamefont {Li}, \citenamefont
  {Zhang}, \citenamefont {Luo}, \citenamefont {Dai},\ and\ \citenamefont
  {Li}}]{Ren2015}%
  \BibitemOpen
  \bibfield  {author} {\bibinfo {author} {\bibfnamefont {X.}~\bibnamefont
  {Ren}}, \bibinfo {author} {\bibfnamefont {L.}~\bibnamefont {Duan}}, \bibinfo
  {author} {\bibfnamefont {Y.}~\bibnamefont {Hu}}, \bibinfo {author}
  {\bibfnamefont {J.}~\bibnamefont {Li}}, \bibinfo {author} {\bibfnamefont
  {R.}~\bibnamefont {Zhang}}, \bibinfo {author} {\bibfnamefont
  {H.}~\bibnamefont {Luo}}, \bibinfo {author} {\bibfnamefont {P.}~\bibnamefont
  {Dai}}, \ and\ \bibinfo {author} {\bibfnamefont {Y.}~\bibnamefont {Li}},\
  }\href {\doibase 10.1103/PhysRevLett.115.197002} {\bibfield  {journal}
  {\bibinfo  {journal} {Phys. Rev. Lett.}\ }\textbf {\bibinfo {volume} {115}},\
  \bibinfo {pages} {197002} (\bibinfo {year} {2015})}\BibitemShut {NoStop}%
\bibitem [{\citenamefont {Dioguardi}\ \emph {et~al.}(2013)\citenamefont
  {Dioguardi}, \citenamefont {Crocker}, \citenamefont {Shockley}, \citenamefont
  {Lin}, \citenamefont {Shirer}, \citenamefont {Nisson}, \citenamefont
  {Lawson}, \citenamefont {apRoberts Warren}, \citenamefont {Canfield},
  \citenamefont {Bud'ko}, \citenamefont {Ran},\ and\ \citenamefont
  {Curro}}]{Dioguardi2013}%
  \BibitemOpen
  \bibfield  {author} {\bibinfo {author} {\bibfnamefont {A.~P.}\ \bibnamefont
  {Dioguardi}}, \bibinfo {author} {\bibfnamefont {J.}~\bibnamefont {Crocker}},
  \bibinfo {author} {\bibfnamefont {A.~C.}\ \bibnamefont {Shockley}}, \bibinfo
  {author} {\bibfnamefont {C.~H.}\ \bibnamefont {Lin}}, \bibinfo {author}
  {\bibfnamefont {K.~R.}\ \bibnamefont {Shirer}}, \bibinfo {author}
  {\bibfnamefont {D.~M.}\ \bibnamefont {Nisson}}, \bibinfo {author}
  {\bibfnamefont {M.~M.}\ \bibnamefont {Lawson}}, \bibinfo {author}
  {\bibfnamefont {N.}~\bibnamefont {apRoberts Warren}}, \bibinfo {author}
  {\bibfnamefont {P.~C.}\ \bibnamefont {Canfield}}, \bibinfo {author}
  {\bibfnamefont {S.~L.}\ \bibnamefont {Bud'ko}}, \bibinfo {author}
  {\bibfnamefont {S.}~\bibnamefont {Ran}}, \ and\ \bibinfo {author}
  {\bibfnamefont {N.~J.}\ \bibnamefont {Curro}},\ }\href {\doibase
  10.1103/PhysRevLett.111.207201} {\bibfield  {journal} {\bibinfo  {journal}
  {Phys. Rev. Lett.}\ }\textbf {\bibinfo {volume} {111}},\ \bibinfo {pages}
  {207201} (\bibinfo {year} {2013})}\BibitemShut {NoStop}%
\bibitem [{\citenamefont {Ning}\ \emph {et~al.}(2010)\citenamefont {Ning},
  \citenamefont {Ahilan}, \citenamefont {Imai}, \citenamefont {Sefat},
  \citenamefont {McGuire}, \citenamefont {Sales}, \citenamefont {Mandrus},
  \citenamefont {Cheng}, \citenamefont {Shen},\ and\ \citenamefont
  {Wen}}]{Ning2010}%
  \BibitemOpen
  \bibfield  {author} {\bibinfo {author} {\bibfnamefont {F.~L.}\ \bibnamefont
  {Ning}}, \bibinfo {author} {\bibfnamefont {K.}~\bibnamefont {Ahilan}},
  \bibinfo {author} {\bibfnamefont {T.}~\bibnamefont {Imai}}, \bibinfo {author}
  {\bibfnamefont {A.~S.}\ \bibnamefont {Sefat}}, \bibinfo {author}
  {\bibfnamefont {M.~A.}\ \bibnamefont {McGuire}}, \bibinfo {author}
  {\bibfnamefont {B.~C.}\ \bibnamefont {Sales}}, \bibinfo {author}
  {\bibfnamefont {D.}~\bibnamefont {Mandrus}}, \bibinfo {author} {\bibfnamefont
  {P.}~\bibnamefont {Cheng}}, \bibinfo {author} {\bibfnamefont
  {B.}~\bibnamefont {Shen}}, \ and\ \bibinfo {author} {\bibfnamefont {H.-H.}\
  \bibnamefont {Wen}},\ }\href {\doibase 10.1103/PhysRevLett.104.037001}
  {\bibfield  {journal} {\bibinfo  {journal} {Phys. Rev. Lett.}\ }\textbf
  {\bibinfo {volume} {104}},\ \bibinfo {pages} {037001} (\bibinfo {year}
  {2010})}\BibitemShut {NoStop}%
\bibitem [{\citenamefont {Kasahara}\ \emph {et~al.}(2012)\citenamefont
  {Kasahara}, \citenamefont {Shi}, \citenamefont {Hashimoto}, \citenamefont
  {Tonegawa}, \citenamefont {Mizukami}, \citenamefont {Shibauchi},
  \citenamefont {Sugimoto}, \citenamefont {Fukuda}, \citenamefont {Terashima},
  \citenamefont {Nevidomskyy},\ and\ \citenamefont {Matsuda}}]{Kasahara2012}%
  \BibitemOpen
  \bibfield  {author} {\bibinfo {author} {\bibfnamefont {S.}~\bibnamefont
  {Kasahara}}, \bibinfo {author} {\bibfnamefont {H.~J.}\ \bibnamefont {Shi}},
  \bibinfo {author} {\bibfnamefont {K.}~\bibnamefont {Hashimoto}}, \bibinfo
  {author} {\bibfnamefont {S.}~\bibnamefont {Tonegawa}}, \bibinfo {author}
  {\bibfnamefont {Y.}~\bibnamefont {Mizukami}}, \bibinfo {author}
  {\bibfnamefont {T.}~\bibnamefont {Shibauchi}}, \bibinfo {author}
  {\bibfnamefont {K.}~\bibnamefont {Sugimoto}}, \bibinfo {author}
  {\bibfnamefont {T.}~\bibnamefont {Fukuda}}, \bibinfo {author} {\bibfnamefont
  {T.}~\bibnamefont {Terashima}}, \bibinfo {author} {\bibfnamefont {A.~H.}\
  \bibnamefont {Nevidomskyy}}, \ and\ \bibinfo {author} {\bibfnamefont
  {Y.}~\bibnamefont {Matsuda}},\ }\href {\doibase 10.1038/nature11178}
  {\bibfield  {journal} {\bibinfo  {journal} {Nat. Lett.}\ }\textbf {\bibinfo
  {volume} {486}},\ \bibinfo {pages} {382} (\bibinfo {year}
  {2012})}\BibitemShut {NoStop}%
\bibitem [{\citenamefont {Xu}\ \emph {et~al.}(2014)\citenamefont {Xu},
  \citenamefont {Jiao}, \citenamefont {Zhou}, \citenamefont {Li}, \citenamefont
  {Chen}, \citenamefont {Cao}, \citenamefont {Dai}, \citenamefont {Bangura},\
  and\ \citenamefont {Cao}}]{Xu2014}%
  \BibitemOpen
  \bibfield  {author} {\bibinfo {author} {\bibfnamefont {X.}~\bibnamefont
  {Xu}}, \bibinfo {author} {\bibfnamefont {W.~H.}\ \bibnamefont {Jiao}},
  \bibinfo {author} {\bibfnamefont {N.}~\bibnamefont {Zhou}}, \bibinfo {author}
  {\bibfnamefont {Y.~K.}\ \bibnamefont {Li}}, \bibinfo {author} {\bibfnamefont
  {B.}~\bibnamefont {Chen}}, \bibinfo {author} {\bibfnamefont {C.}~\bibnamefont
  {Cao}}, \bibinfo {author} {\bibfnamefont {J.}~\bibnamefont {Dai}}, \bibinfo
  {author} {\bibfnamefont {A.~F.}\ \bibnamefont {Bangura}}, \ and\ \bibinfo
  {author} {\bibfnamefont {G.}~\bibnamefont {Cao}},\ }\href {\doibase
  10.1103/PhysRevB.89.104517} {\bibfield  {journal} {\bibinfo  {journal} {Phys.
  Rev. B}\ }\textbf {\bibinfo {volume} {89}},\ \bibinfo {pages} {104517}
  (\bibinfo {year} {2014})}\BibitemShut {NoStop}%
\bibitem [{\citenamefont {Kim}\ \emph {et~al.}(2013)\citenamefont {Kim},
  \citenamefont {Jung}, \citenamefont {Han}, \citenamefont {Choi},
  \citenamefont {Chen}, \citenamefont {Devereaux}, \citenamefont {Chainani},
  \citenamefont {Miyawaki}, \citenamefont {Takata}, \citenamefont {Tanaka},
  \citenamefont {Oura}, \citenamefont {Shin}, \citenamefont {Singh},
  \citenamefont {Lee}, \citenamefont {Kim},\ and\ \citenamefont
  {Kim}}]{Kim2013}%
  \BibitemOpen
  \bibfield  {author} {\bibinfo {author} {\bibfnamefont {Y.~K.}\ \bibnamefont
  {Kim}}, \bibinfo {author} {\bibfnamefont {W.~S.}\ \bibnamefont {Jung}},
  \bibinfo {author} {\bibfnamefont {G.~R.}\ \bibnamefont {Han}}, \bibinfo
  {author} {\bibfnamefont {K.-Y.}\ \bibnamefont {Choi}}, \bibinfo {author}
  {\bibfnamefont {C.-C.}\ \bibnamefont {Chen}}, \bibinfo {author}
  {\bibfnamefont {T.~P.}\ \bibnamefont {Devereaux}}, \bibinfo {author}
  {\bibfnamefont {A.}~\bibnamefont {Chainani}}, \bibinfo {author}
  {\bibfnamefont {J.}~\bibnamefont {Miyawaki}}, \bibinfo {author}
  {\bibfnamefont {Y.}~\bibnamefont {Takata}}, \bibinfo {author} {\bibfnamefont
  {Y.}~\bibnamefont {Tanaka}}, \bibinfo {author} {\bibfnamefont
  {M.}~\bibnamefont {Oura}}, \bibinfo {author} {\bibfnamefont {S.}~\bibnamefont
  {Shin}}, \bibinfo {author} {\bibfnamefont {A.~P.}\ \bibnamefont {Singh}},
  \bibinfo {author} {\bibfnamefont {H.~G.}\ \bibnamefont {Lee}}, \bibinfo
  {author} {\bibfnamefont {J.-Y.}\ \bibnamefont {Kim}}, \ and\ \bibinfo
  {author} {\bibfnamefont {C.}~\bibnamefont {Kim}},\ }\href {\doibase
  10.1103/PhysRevLett.111.217001} {\bibfield  {journal} {\bibinfo  {journal}
  {Phys. Rev. Lett.}\ }\textbf {\bibinfo {volume} {111}},\ \bibinfo {pages}
  {217001} (\bibinfo {year} {2013})}\BibitemShut {NoStop}%
\bibitem [{\citenamefont {Arham}\ and\ \citenamefont
  {Greene}(2013)}]{Arham2013}%
  \BibitemOpen
  \bibfield  {author} {\bibinfo {author} {\bibfnamefont {H.~Z.}\ \bibnamefont
  {Arham}}\ and\ \bibinfo {author} {\bibfnamefont {L.~H.}\ \bibnamefont
  {Greene}},\ }\href {\doibase 10.1016/j.cossms.2013.06.001} {\bibfield
  {journal} {\bibinfo  {journal} {Current Opinion in Solid State Mat Sci}\
  }\textbf {\bibinfo {volume} {17}},\ \bibinfo {pages} {81} (\bibinfo {year}
  {2013})},\ \bibinfo {note} {fe-based Superconductors}\BibitemShut {NoStop}%
\bibitem [{\citenamefont {Yi}\ \emph {et~al.}(2011)\citenamefont {Yi},
  \citenamefont {Lu}, \citenamefont {Chu}, \citenamefont {Analytis},
  \citenamefont {Sorini}, \citenamefont {Kemper}, \citenamefont {Moritz},
  \citenamefont {Mo}, \citenamefont {Moore}, \citenamefont {Hashimoto},
  \citenamefont {Lee}, \citenamefont {Hussain}, \citenamefont {Devereaux},
  \citenamefont {Fisher},\ and\ \citenamefont {Shen}}]{Yi2011}%
  \BibitemOpen
  \bibfield  {author} {\bibinfo {author} {\bibfnamefont {M.}~\bibnamefont
  {Yi}}, \bibinfo {author} {\bibfnamefont {D.}~\bibnamefont {Lu}}, \bibinfo
  {author} {\bibfnamefont {J.-H.}\ \bibnamefont {Chu}}, \bibinfo {author}
  {\bibfnamefont {J.~G.}\ \bibnamefont {Analytis}}, \bibinfo {author}
  {\bibfnamefont {A.~P.}\ \bibnamefont {Sorini}}, \bibinfo {author}
  {\bibfnamefont {A.~F.}\ \bibnamefont {Kemper}}, \bibinfo {author}
  {\bibfnamefont {B.}~\bibnamefont {Moritz}}, \bibinfo {author} {\bibfnamefont
  {S.-K.}\ \bibnamefont {Mo}}, \bibinfo {author} {\bibfnamefont {R.~G.}\
  \bibnamefont {Moore}}, \bibinfo {author} {\bibfnamefont {M.}~\bibnamefont
  {Hashimoto}}, \bibinfo {author} {\bibfnamefont {W.-S.}\ \bibnamefont {Lee}},
  \bibinfo {author} {\bibfnamefont {Z.}~\bibnamefont {Hussain}}, \bibinfo
  {author} {\bibfnamefont {T.~P.}\ \bibnamefont {Devereaux}}, \bibinfo {author}
  {\bibfnamefont {I.~R.}\ \bibnamefont {Fisher}}, \ and\ \bibinfo {author}
  {\bibfnamefont {Z.-X.}\ \bibnamefont {Shen}},\ }\href {\doibase
  10.1073/pnas.1015572108} {\bibfield  {journal} {\bibinfo  {journal} {Proc.
  Nat. Acad. Sciences}\ }\textbf {\bibinfo {volume} {108}},\ \bibinfo {pages}
  {6878} (\bibinfo {year} {2011})}\BibitemShut {NoStop}%
\bibitem [{\citenamefont {Takagi}\ \emph {et~al.}(1992)\citenamefont {Takagi},
  \citenamefont {Batlogg}, \citenamefont {Kao}, \citenamefont {Kwo},
  \citenamefont {Cava}, \citenamefont {Krajewski},\ and\ \citenamefont
  {Peck}}]{Takagi1992}%
  \BibitemOpen
  \bibfield  {author} {\bibinfo {author} {\bibfnamefont {H.}~\bibnamefont
  {Takagi}}, \bibinfo {author} {\bibfnamefont {B.}~\bibnamefont {Batlogg}},
  \bibinfo {author} {\bibfnamefont {H.~L.}\ \bibnamefont {Kao}}, \bibinfo
  {author} {\bibfnamefont {J.}~\bibnamefont {Kwo}}, \bibinfo {author}
  {\bibfnamefont {R.~J.}\ \bibnamefont {Cava}}, \bibinfo {author}
  {\bibfnamefont {J.~J.}\ \bibnamefont {Krajewski}}, \ and\ \bibinfo {author}
  {\bibfnamefont {W.~F.}\ \bibnamefont {Peck}},\ }\href {\doibase
  10.1103/PhysRevLett.69.2975} {\bibfield  {journal} {\bibinfo  {journal}
  {Phys. Rev. Lett.}\ }\textbf {\bibinfo {volume} {69}},\ \bibinfo {pages}
  {2975} (\bibinfo {year} {1992})}\BibitemShut {NoStop}%
\bibitem [{\citenamefont {Ando}\ \emph {et~al.}(2004)\citenamefont {Ando},
  \citenamefont {Komiya}, \citenamefont {Segawa}, \citenamefont {Ono},\ and\
  \citenamefont {Kurita}}]{Ando2004}%
  \BibitemOpen
  \bibfield  {author} {\bibinfo {author} {\bibfnamefont {Y.}~\bibnamefont
  {Ando}}, \bibinfo {author} {\bibfnamefont {S.}~\bibnamefont {Komiya}},
  \bibinfo {author} {\bibfnamefont {K.}~\bibnamefont {Segawa}}, \bibinfo
  {author} {\bibfnamefont {S.}~\bibnamefont {Ono}}, \ and\ \bibinfo {author}
  {\bibfnamefont {Y.}~\bibnamefont {Kurita}},\ }\href {\doibase
  10.1103/PhysRevLett.93.267001} {\bibfield  {journal} {\bibinfo  {journal}
  {Phys. Rev. Lett.}\ }\textbf {\bibinfo {volume} {93}},\ \bibinfo {pages}
  {267001} (\bibinfo {year} {2004})}\BibitemShut {NoStop}%
\bibitem [{\citenamefont {Hess}\ \emph {et~al.}(2009)\citenamefont {Hess},
  \citenamefont {Kondrat}, \citenamefont {Narduzzo}, \citenamefont
  {Hamann-Borrero}, \citenamefont {Klingeler}, \citenamefont {Werner},
  \citenamefont {Behr},\ and\ \citenamefont {Büchner}}]{Hess2009}%
  \BibitemOpen
  \bibfield  {author} {\bibinfo {author} {\bibfnamefont {C.}~\bibnamefont
  {Hess}}, \bibinfo {author} {\bibfnamefont {A.}~\bibnamefont {Kondrat}},
  \bibinfo {author} {\bibfnamefont {A.}~\bibnamefont {Narduzzo}}, \bibinfo
  {author} {\bibfnamefont {J.~E.}\ \bibnamefont {Hamann-Borrero}}, \bibinfo
  {author} {\bibfnamefont {R.}~\bibnamefont {Klingeler}}, \bibinfo {author}
  {\bibfnamefont {J.}~\bibnamefont {Werner}}, \bibinfo {author} {\bibfnamefont
  {G.}~\bibnamefont {Behr}}, \ and\ \bibinfo {author} {\bibfnamefont
  {B.}~\bibnamefont {Büchner}},\ }\href
  {http://stacks.iop.org/0295-5075/87/i=1/a=17005} {\bibfield  {journal}
  {\bibinfo  {journal} {EPL}\ }\textbf {\bibinfo {volume} {87}},\ \bibinfo
  {pages} {17005} (\bibinfo {year} {2009})}\BibitemShut {NoStop}%
\bibitem [{\citenamefont {Yan}\ \emph {et~al.}(2013)\citenamefont {Yan},
  \citenamefont {Wang}, \citenamefont {Luo}, \citenamefont {Sun}, \citenamefont
  {Ying}, \citenamefont {Ye}, \citenamefont {Chen}, \citenamefont {Ma},\ and\
  \citenamefont {Chen}}]{Yan2013}%
  \BibitemOpen
  \bibfield  {author} {\bibinfo {author} {\bibfnamefont {Y.~J.}\ \bibnamefont
  {Yan}}, \bibinfo {author} {\bibfnamefont {A.~F.}\ \bibnamefont {Wang}},
  \bibinfo {author} {\bibfnamefont {X.~G.}\ \bibnamefont {Luo}}, \bibinfo
  {author} {\bibfnamefont {Z.}~\bibnamefont {Sun}}, \bibinfo {author}
  {\bibfnamefont {J.~J.}\ \bibnamefont {Ying}}, \bibinfo {author}
  {\bibfnamefont {G.~J.}\ \bibnamefont {Ye}}, \bibinfo {author} {\bibfnamefont
  {P.}~\bibnamefont {Chen}}, \bibinfo {author} {\bibfnamefont {J.~Q.}\
  \bibnamefont {Ma}}, \ and\ \bibinfo {author} {\bibfnamefont {X.~H.}\
  \bibnamefont {Chen}},\ }\href {http://arxiv.org/abs/1301.1734} {\bibfield
  {journal} {\bibinfo  {journal} {ArXiv e-prints}\ } (\bibinfo {year}
  {2013})},\ \Eprint {http://arxiv.org/abs/1301.1734} {arXiv:1301.1734}
  \BibitemShut {NoStop}%
\bibitem [{\citenamefont {Presniakov}\ \emph {et~al.}(2013)\citenamefont
  {Presniakov}, \citenamefont {Morozov}, \citenamefont {Sobolev}, \citenamefont
  {Roslova}, \citenamefont {Boltalin}, \citenamefont {Son}, \citenamefont
  {Volkova}, \citenamefont {Vasiliev}, \citenamefont {Wurmehl},\ and\
  \citenamefont {B\"uchner}}]{Presniakov2013}%
  \BibitemOpen
  \bibfield  {author} {\bibinfo {author} {\bibfnamefont {I.}~\bibnamefont
  {Presniakov}}, \bibinfo {author} {\bibfnamefont {I.}~\bibnamefont {Morozov}},
  \bibinfo {author} {\bibfnamefont {A.}~\bibnamefont {Sobolev}}, \bibinfo
  {author} {\bibfnamefont {M.}~\bibnamefont {Roslova}}, \bibinfo {author}
  {\bibfnamefont {A.}~\bibnamefont {Boltalin}}, \bibinfo {author}
  {\bibfnamefont {V.}~\bibnamefont {Son}}, \bibinfo {author} {\bibfnamefont
  {O.}~\bibnamefont {Volkova}}, \bibinfo {author} {\bibfnamefont
  {A.}~\bibnamefont {Vasiliev}}, \bibinfo {author} {\bibfnamefont
  {S.}~\bibnamefont {Wurmehl}}, \ and\ \bibinfo {author} {\bibfnamefont
  {B.}~\bibnamefont {B\"uchner}},\ }\href
  {http://stacks.iop.org/0953-8984/25/i=34/a=346003} {\bibfield  {journal}
  {\bibinfo  {journal} {J. Phys.: Cond. Mat.}\ }\textbf {\bibinfo {volume}
  {25}},\ \bibinfo {pages} {346003} (\bibinfo {year} {2013})}\BibitemShut
  {NoStop}%
\bibitem [{\citenamefont {Wang}\ \emph {et~al.}(2012)\citenamefont {Wang},
  \citenamefont {Luo}, \citenamefont {Yan}, \citenamefont {Ying}, \citenamefont
  {Xiang}, \citenamefont {Ye}, \citenamefont {Cheng}, \citenamefont {Li},
  \citenamefont {Hu},\ and\ \citenamefont {Chen}}]{Wang2012}%
  \BibitemOpen
  \bibfield  {author} {\bibinfo {author} {\bibfnamefont {A.~F.}\ \bibnamefont
  {Wang}}, \bibinfo {author} {\bibfnamefont {X.~G.}\ \bibnamefont {Luo}},
  \bibinfo {author} {\bibfnamefont {Y.~J.}\ \bibnamefont {Yan}}, \bibinfo
  {author} {\bibfnamefont {J.~J.}\ \bibnamefont {Ying}}, \bibinfo {author}
  {\bibfnamefont {Z.~J.}\ \bibnamefont {Xiang}}, \bibinfo {author}
  {\bibfnamefont {G.~J.}\ \bibnamefont {Ye}}, \bibinfo {author} {\bibfnamefont
  {P.}~\bibnamefont {Cheng}}, \bibinfo {author} {\bibfnamefont {Z.~Y.}\
  \bibnamefont {Li}}, \bibinfo {author} {\bibfnamefont {W.~J.}\ \bibnamefont
  {Hu}}, \ and\ \bibinfo {author} {\bibfnamefont {X.~H.}\ \bibnamefont
  {Chen}},\ }\href {\doibase 10.1103/PhysRevB.85.224521} {\bibfield  {journal}
  {\bibinfo  {journal} {Phys. Rev. B}\ }\textbf {\bibinfo {volume} {85}},\
  \bibinfo {pages} {224521} (\bibinfo {year} {2012})}\BibitemShut {NoStop}%
\bibitem [{\citenamefont {Ziman}(2007)}]{Ziman2007}%
  \BibitemOpen
  \bibfield  {author} {\bibinfo {author} {\bibfnamefont {J.~M.}\ \bibnamefont
  {Ziman}},\ }\href@noop {} {\emph {\bibinfo {title} {Electrons and Phonons}}}\
  (\bibinfo  {publisher} {Oxford: Clarendon Press},\ \bibinfo {year}
  {2007})\BibitemShut {NoStop}%
\bibitem [{\citenamefont {Spyrison}\ \emph {et~al.}(2012)\citenamefont
  {Spyrison}, \citenamefont {Tanatar}, \citenamefont {Cho}, \citenamefont
  {Song}, \citenamefont {Dai}, \citenamefont {Zhang},\ and\ \citenamefont
  {Prozorov}}]{Spyrison2012}%
  \BibitemOpen
  \bibfield  {author} {\bibinfo {author} {\bibfnamefont {N.}~\bibnamefont
  {Spyrison}}, \bibinfo {author} {\bibfnamefont {M.~A.}\ \bibnamefont
  {Tanatar}}, \bibinfo {author} {\bibfnamefont {K.}~\bibnamefont {Cho}},
  \bibinfo {author} {\bibfnamefont {Y.}~\bibnamefont {Song}}, \bibinfo {author}
  {\bibfnamefont {P.}~\bibnamefont {Dai}}, \bibinfo {author} {\bibfnamefont
  {C.}~\bibnamefont {Zhang}}, \ and\ \bibinfo {author} {\bibfnamefont
  {R.}~\bibnamefont {Prozorov}},\ }\href {\doibase 10.1103/PhysRevB.86.144528}
  {\bibfield  {journal} {\bibinfo  {journal} {Phys. Rev. B}\ }\textbf {\bibinfo
  {volume} {86}},\ \bibinfo {pages} {144528} (\bibinfo {year}
  {2012})}\BibitemShut {NoStop}%
\bibitem [{\citenamefont {Deng}\ \emph
  {et~al.}(2015{\natexlab{b}})\citenamefont {Deng}, \citenamefont {Xing},
  \citenamefont {Liu}, \citenamefont {Yang},\ and\ \citenamefont
  {Wen}}]{Deng2015a}%
  \BibitemOpen
  \bibfield  {author} {\bibinfo {author} {\bibfnamefont {Q.}~\bibnamefont
  {Deng}}, \bibinfo {author} {\bibfnamefont {J.}~\bibnamefont {Xing}}, \bibinfo
  {author} {\bibfnamefont {J.}~\bibnamefont {Liu}}, \bibinfo {author}
  {\bibfnamefont {H.}~\bibnamefont {Yang}}, \ and\ \bibinfo {author}
  {\bibfnamefont {H.-H.}\ \bibnamefont {Wen}},\ }\href {\doibase
  10.1103/PhysRevB.92.014510} {\bibfield  {journal} {\bibinfo  {journal} {Phys.
  Rev. B}\ }\textbf {\bibinfo {volume} {92}},\ \bibinfo {pages} {014510}
  (\bibinfo {year} {2015}{\natexlab{b}})}\BibitemShut {NoStop}%
\bibitem [{\citenamefont {He}\ \emph {et~al.}(2010)\citenamefont {He},
  \citenamefont {Zhang}, \citenamefont {Xie}, \citenamefont {Wang},
  \citenamefont {Yang}, \citenamefont {Zhou}, \citenamefont {Chen},
  \citenamefont {Arita}, \citenamefont {Shimada}, \citenamefont {Namatame},
  \citenamefont {Taniguchi}, \citenamefont {Chen}, \citenamefont {Hu},\ and\
  \citenamefont {Feng}}]{He2010}%
  \BibitemOpen
  \bibfield  {author} {\bibinfo {author} {\bibfnamefont {C.}~\bibnamefont
  {He}}, \bibinfo {author} {\bibfnamefont {Y.}~\bibnamefont {Zhang}}, \bibinfo
  {author} {\bibfnamefont {B.~P.}\ \bibnamefont {Xie}}, \bibinfo {author}
  {\bibfnamefont {X.~F.}\ \bibnamefont {Wang}}, \bibinfo {author}
  {\bibfnamefont {L.~X.}\ \bibnamefont {Yang}}, \bibinfo {author}
  {\bibfnamefont {B.}~\bibnamefont {Zhou}}, \bibinfo {author} {\bibfnamefont
  {F.}~\bibnamefont {Chen}}, \bibinfo {author} {\bibfnamefont {M.}~\bibnamefont
  {Arita}}, \bibinfo {author} {\bibfnamefont {K.}~\bibnamefont {Shimada}},
  \bibinfo {author} {\bibfnamefont {H.}~\bibnamefont {Namatame}}, \bibinfo
  {author} {\bibfnamefont {M.}~\bibnamefont {Taniguchi}}, \bibinfo {author}
  {\bibfnamefont {X.~H.}\ \bibnamefont {Chen}}, \bibinfo {author}
  {\bibfnamefont {J.~P.}\ \bibnamefont {Hu}}, \ and\ \bibinfo {author}
  {\bibfnamefont {D.~L.}\ \bibnamefont {Feng}},\ }\href {\doibase
  10.1103/PhysRevLett.105.117002} {\bibfield  {journal} {\bibinfo  {journal}
  {Phys. Rev. Lett.}\ }\textbf {\bibinfo {volume} {105}},\ \bibinfo {pages}
  {117002} (\bibinfo {year} {2010})}\BibitemShut {NoStop}%
\bibitem [{\citenamefont {Chen}\ \emph {et~al.}(2009)\citenamefont {Chen},
  \citenamefont {Hu}, \citenamefont {Luo},\ and\ \citenamefont
  {Wang}}]{Chen2009}%
  \BibitemOpen
  \bibfield  {author} {\bibinfo {author} {\bibfnamefont {G.~F.}\ \bibnamefont
  {Chen}}, \bibinfo {author} {\bibfnamefont {W.~Z.}\ \bibnamefont {Hu}},
  \bibinfo {author} {\bibfnamefont {J.~L.}\ \bibnamefont {Luo}}, \ and\
  \bibinfo {author} {\bibfnamefont {N.~L.}\ \bibnamefont {Wang}},\ }\href
  {\doibase 10.1103/PhysRevLett.102.227004} {\bibfield  {journal} {\bibinfo
  {journal} {Phys. Rev. Lett.}\ }\textbf {\bibinfo {volume} {102}},\ \bibinfo
  {pages} {227004} (\bibinfo {year} {2009})}\BibitemShut {NoStop}%
\bibitem [{\citenamefont {Wang}\ \emph
  {et~al.}(2013{\natexlab{a}})\citenamefont {Wang}, \citenamefont {Ying},
  \citenamefont {Luo}, \citenamefont {Yan}, \citenamefont {Liu}, \citenamefont
  {Xiang}, \citenamefont {Cheng}, \citenamefont {Ye}, \citenamefont {Zou},
  \citenamefont {Sun},\ and\ \citenamefont {Chen}}]{Wang2013a}%
  \BibitemOpen
  \bibfield  {author} {\bibinfo {author} {\bibfnamefont {A.~F.}\ \bibnamefont
  {Wang}}, \bibinfo {author} {\bibfnamefont {J.~J.}\ \bibnamefont {Ying}},
  \bibinfo {author} {\bibfnamefont {X.~G.}\ \bibnamefont {Luo}}, \bibinfo
  {author} {\bibfnamefont {Y.~J.}\ \bibnamefont {Yan}}, \bibinfo {author}
  {\bibfnamefont {D.~Y.}\ \bibnamefont {Liu}}, \bibinfo {author} {\bibfnamefont
  {Z.~J.}\ \bibnamefont {Xiang}}, \bibinfo {author} {\bibfnamefont
  {P.}~\bibnamefont {Cheng}}, \bibinfo {author} {\bibfnamefont {G.~J.}\
  \bibnamefont {Ye}}, \bibinfo {author} {\bibfnamefont {L.~J.}\ \bibnamefont
  {Zou}}, \bibinfo {author} {\bibfnamefont {Z.}~\bibnamefont {Sun}}, \ and\
  \bibinfo {author} {\bibfnamefont {X.~H.}\ \bibnamefont {Chen}},\ }\href
  {http://stacks.iop.org/1367-2630/15/i=4/a=043048} {\bibfield  {journal}
  {\bibinfo  {journal} {New J. Phys.}\ }\textbf {\bibinfo {volume} {15}},\
  \bibinfo {pages} {043048} (\bibinfo {year} {2013}{\natexlab{a}})}\BibitemShut
  {NoStop}%
\bibitem [{\citenamefont {Wang}\ \emph
  {et~al.}(2013{\natexlab{b}})\citenamefont {Wang}, \citenamefont {Wang},
  \citenamefont {Sou}, \citenamefont {Yang}, \citenamefont {Chang},
  \citenamefont {Redding}, \citenamefont {Song}, \citenamefont {Dai},\ and\
  \citenamefont {Zhang}}]{Wang2013}%
  \BibitemOpen
  \bibfield  {author} {\bibinfo {author} {\bibfnamefont {L.~M.}\ \bibnamefont
  {Wang}}, \bibinfo {author} {\bibfnamefont {C.-Y.}\ \bibnamefont {Wang}},
  \bibinfo {author} {\bibfnamefont {U.-C.}\ \bibnamefont {Sou}}, \bibinfo
  {author} {\bibfnamefont {H.~C.}\ \bibnamefont {Yang}}, \bibinfo {author}
  {\bibfnamefont {L.~J.}\ \bibnamefont {Chang}}, \bibinfo {author}
  {\bibfnamefont {C.}~\bibnamefont {Redding}}, \bibinfo {author} {\bibfnamefont
  {Y.}~\bibnamefont {Song}}, \bibinfo {author} {\bibfnamefont {P.}~\bibnamefont
  {Dai}}, \ and\ \bibinfo {author} {\bibfnamefont {C.}~\bibnamefont {Zhang}},\
  }\href {\doibase 10.1088/0953-8984/25/39/395702} {\bibfield  {journal}
  {\bibinfo  {journal} {J. Phys.: Cond. Mat.}\ }\textbf {\bibinfo {volume}
  {25}},\ \bibinfo {pages} {395702} (\bibinfo {year}
  {2013}{\natexlab{b}})}\BibitemShut {NoStop}%
\bibitem [{\citenamefont {Chien}\ \emph {et~al.}(1991)\citenamefont {Chien},
  \citenamefont {Wang},\ and\ \citenamefont {Ong}}]{Chien1991}%
  \BibitemOpen
  \bibfield  {author} {\bibinfo {author} {\bibfnamefont {T.~R.}\ \bibnamefont
  {Chien}}, \bibinfo {author} {\bibfnamefont {Z.~Z.}\ \bibnamefont {Wang}}, \
  and\ \bibinfo {author} {\bibfnamefont {N.~P.}\ \bibnamefont {Ong}},\ }\href
  {\doibase 10.1103/PhysRevLett.67.2088} {\bibfield  {journal} {\bibinfo
  {journal} {Phys. Rev. Lett.}\ }\textbf {\bibinfo {volume} {67}},\ \bibinfo
  {pages} {2088} (\bibinfo {year} {1991})}\BibitemShut {NoStop}%
\bibitem [{\citenamefont {Anderson}(1991)}]{Anderson1991}%
  \BibitemOpen
  \bibfield  {author} {\bibinfo {author} {\bibfnamefont {P.~W.}\ \bibnamefont
  {Anderson}},\ }\href {\doibase 10.1103/PhysRevLett.67.2092} {\bibfield
  {journal} {\bibinfo  {journal} {Phys. Rev. Lett.}\ }\textbf {\bibinfo
  {volume} {67}},\ \bibinfo {pages} {2092} (\bibinfo {year}
  {1991})}\BibitemShut {NoStop}%
\bibitem [{\citenamefont {Arushanov}\ \emph {et~al.}(2011)\citenamefont
  {Arushanov}, \citenamefont {Levcenko}, \citenamefont {Fuchs}, \citenamefont
  {Holzapfel}, \citenamefont {Drechsler},\ and\ \citenamefont
  {Schultz}}]{Arushanov2011}%
  \BibitemOpen
  \bibfield  {author} {\bibinfo {author} {\bibfnamefont {E.}~\bibnamefont
  {Arushanov}}, \bibinfo {author} {\bibfnamefont {S.}~\bibnamefont {Levcenko}},
  \bibinfo {author} {\bibfnamefont {G.}~\bibnamefont {Fuchs}}, \bibinfo
  {author} {\bibfnamefont {B.}~\bibnamefont {Holzapfel}}, \bibinfo {author}
  {\bibfnamefont {S.}~\bibnamefont {Drechsler}}, \ and\ \bibinfo {author}
  {\bibfnamefont {L.}~\bibnamefont {Schultz}},\ }\href {\doibase
  10.1007/s10948-011-1198-1} {\bibfield  {journal} {\bibinfo  {journal} {J.
  Supercond. Nov. Magn.}\ }\textbf {\bibinfo {volume} {24}},\ \bibinfo {pages}
  {2285} (\bibinfo {year} {2011})}\BibitemShut {NoStop}%
\bibitem [{\citenamefont {B\"ohmer}\ \emph {et~al.}(2014)\citenamefont
  {B\"ohmer}, \citenamefont {Burger}, \citenamefont {Hardy}, \citenamefont
  {Wolf}, \citenamefont {Schweiss}, \citenamefont {Fromknecht}, \citenamefont
  {Reinecker}, \citenamefont {Schranz},\ and\ \citenamefont
  {Meingast}}]{Bohmer2014}%
  \BibitemOpen
  \bibfield  {author} {\bibinfo {author} {\bibfnamefont {A.~E.}\ \bibnamefont
  {B\"ohmer}}, \bibinfo {author} {\bibfnamefont {P.}~\bibnamefont {Burger}},
  \bibinfo {author} {\bibfnamefont {F.}~\bibnamefont {Hardy}}, \bibinfo
  {author} {\bibfnamefont {T.}~\bibnamefont {Wolf}}, \bibinfo {author}
  {\bibfnamefont {P.}~\bibnamefont {Schweiss}}, \bibinfo {author}
  {\bibfnamefont {R.}~\bibnamefont {Fromknecht}}, \bibinfo {author}
  {\bibfnamefont {M.}~\bibnamefont {Reinecker}}, \bibinfo {author}
  {\bibfnamefont {W.}~\bibnamefont {Schranz}}, \ and\ \bibinfo {author}
  {\bibfnamefont {C.}~\bibnamefont {Meingast}},\ }\href {\doibase
  10.1103/PhysRevLett.112.047001} {\bibfield  {journal} {\bibinfo  {journal}
  {Phys. Rev. Lett.}\ }\textbf {\bibinfo {volume} {112}},\ \bibinfo {pages}
  {047001} (\bibinfo {year} {2014})}\BibitemShut {NoStop}%
\bibitem [{\citenamefont {Fernandes}\ \emph {et~al.}(2013)\citenamefont
  {Fernandes}, \citenamefont {B\"ohmer}, \citenamefont {Meingast},\ and\
  \citenamefont {Schmalian}}]{Fernandes2013}%
  \BibitemOpen
  \bibfield  {author} {\bibinfo {author} {\bibfnamefont {R.~M.}\ \bibnamefont
  {Fernandes}}, \bibinfo {author} {\bibfnamefont {A.~E.}\ \bibnamefont
  {B\"ohmer}}, \bibinfo {author} {\bibfnamefont {C.}~\bibnamefont {Meingast}},
  \ and\ \bibinfo {author} {\bibfnamefont {J.}~\bibnamefont {Schmalian}},\
  }\href {\doibase 10.1103/PhysRevLett.111.137001} {\bibfield  {journal}
  {\bibinfo  {journal} {Phys. Rev. Lett.}\ }\textbf {\bibinfo {volume} {111}},\
  \bibinfo {pages} {137001} (\bibinfo {year} {2013})}\BibitemShut {NoStop}%
\bibitem [{\citenamefont {Ishida}\ \emph {et~al.}(2013)\citenamefont {Ishida},
  \citenamefont {Nakajima}, \citenamefont {Liang}, \citenamefont {Kihou},
  \citenamefont {Lee}, \citenamefont {Iyo}, \citenamefont {Eisaki},
  \citenamefont {Kakeshita}, \citenamefont {Tomioka}, \citenamefont {Ito},\
  and\ \citenamefont {Uchida}}]{Ishida2013}%
  \BibitemOpen
  \bibfield  {author} {\bibinfo {author} {\bibfnamefont {S.}~\bibnamefont
  {Ishida}}, \bibinfo {author} {\bibfnamefont {M.}~\bibnamefont {Nakajima}},
  \bibinfo {author} {\bibfnamefont {T.}~\bibnamefont {Liang}}, \bibinfo
  {author} {\bibfnamefont {K.}~\bibnamefont {Kihou}}, \bibinfo {author}
  {\bibfnamefont {C.~H.}\ \bibnamefont {Lee}}, \bibinfo {author} {\bibfnamefont
  {A.}~\bibnamefont {Iyo}}, \bibinfo {author} {\bibfnamefont {H.}~\bibnamefont
  {Eisaki}}, \bibinfo {author} {\bibfnamefont {T.}~\bibnamefont {Kakeshita}},
  \bibinfo {author} {\bibfnamefont {Y.}~\bibnamefont {Tomioka}}, \bibinfo
  {author} {\bibfnamefont {T.}~\bibnamefont {Ito}}, \ and\ \bibinfo {author}
  {\bibfnamefont {S.}~\bibnamefont {Uchida}},\ }\href {\doibase
  10.1103/PhysRevLett.110.207001} {\bibfield  {journal} {\bibinfo  {journal}
  {Phys. Rev. Lett.}\ }\textbf {\bibinfo {volume} {110}},\ \bibinfo {pages}
  {207001} (\bibinfo {year} {2013})}\BibitemShut {NoStop}%
\bibitem [{\citenamefont {Allan}\ \emph {et~al.}(2013)\citenamefont {Allan},
  \citenamefont {Chuang}, \citenamefont {Massee}, \citenamefont {Xie},
  \citenamefont {Ni}, \citenamefont {Bud?ko}, \citenamefont {andQ. Wang},
  \citenamefont {Dessau}, \citenamefont {Canfield}, \citenamefont {Golden},\
  and\ \citenamefont {Davis}}]{Allan2013}%
  \BibitemOpen
  \bibfield  {author} {\bibinfo {author} {\bibfnamefont {M.~P.}\ \bibnamefont
  {Allan}}, \bibinfo {author} {\bibfnamefont {T.-M.}\ \bibnamefont {Chuang}},
  \bibinfo {author} {\bibfnamefont {F.}~\bibnamefont {Massee}}, \bibinfo
  {author} {\bibfnamefont {Y.}~\bibnamefont {Xie}}, \bibinfo {author}
  {\bibfnamefont {N.}~\bibnamefont {Ni}}, \bibinfo {author} {\bibfnamefont
  {S.~L.}\ \bibnamefont {Bud?ko}}, \bibinfo {author} {\bibfnamefont {G.~S.~B.}\
  \bibnamefont {andQ. Wang}}, \bibinfo {author} {\bibfnamefont {D.~S.}\
  \bibnamefont {Dessau}}, \bibinfo {author} {\bibfnamefont {P.~C.}\
  \bibnamefont {Canfield}}, \bibinfo {author} {\bibfnamefont {M.~S.}\
  \bibnamefont {Golden}}, \ and\ \bibinfo {author} {\bibfnamefont {J.~C.}\
  \bibnamefont {Davis}},\ }\href {\doibase 10.1038/nphys2544} {\bibfield
  {journal} {\bibinfo  {journal} {Nat. Phys.}\ }\textbf {\bibinfo {volume}
  {9}},\ \bibinfo {pages} {220} (\bibinfo {year} {2013})}\BibitemShut {NoStop}%
\bibitem [{\citenamefont {Kuo}\ and\ \citenamefont {Fisher}(2014)}]{Kuo2014}%
  \BibitemOpen
  \bibfield  {author} {\bibinfo {author} {\bibfnamefont {H.-H.}\ \bibnamefont
  {Kuo}}\ and\ \bibinfo {author} {\bibfnamefont {I.~R.}\ \bibnamefont
  {Fisher}},\ }\href {\doibase 10.1103/PhysRevLett.112.227001} {\bibfield
  {journal} {\bibinfo  {journal} {Phys. Rev. Lett.}\ }\textbf {\bibinfo
  {volume} {112}},\ \bibinfo {pages} {227001} (\bibinfo {year}
  {2014})}\BibitemShut {NoStop}%
\bibitem [{\citenamefont {Kuo}\ \emph {et~al.}(2011)\citenamefont {Kuo},
  \citenamefont {Chu}, \citenamefont {Riggs}, \citenamefont {Yu}, \citenamefont
  {McMahon}, \citenamefont {De~Greve}, \citenamefont {Yamamoto}, \citenamefont
  {Analytis},\ and\ \citenamefont {Fisher}}]{Kuo2011}%
  \BibitemOpen
  \bibfield  {author} {\bibinfo {author} {\bibfnamefont {H.-H.}\ \bibnamefont
  {Kuo}}, \bibinfo {author} {\bibfnamefont {J.-H.}\ \bibnamefont {Chu}},
  \bibinfo {author} {\bibfnamefont {S.~C.}\ \bibnamefont {Riggs}}, \bibinfo
  {author} {\bibfnamefont {L.}~\bibnamefont {Yu}}, \bibinfo {author}
  {\bibfnamefont {P.~L.}\ \bibnamefont {McMahon}}, \bibinfo {author}
  {\bibfnamefont {K.}~\bibnamefont {De~Greve}}, \bibinfo {author}
  {\bibfnamefont {Y.}~\bibnamefont {Yamamoto}}, \bibinfo {author}
  {\bibfnamefont {J.~G.}\ \bibnamefont {Analytis}}, \ and\ \bibinfo {author}
  {\bibfnamefont {I.~R.}\ \bibnamefont {Fisher}},\ }\href {\doibase
  10.1103/PhysRevB.84.054540} {\bibfield  {journal} {\bibinfo  {journal} {Phys.
  Rev. B}\ }\textbf {\bibinfo {volume} {84}},\ \bibinfo {pages} {054540}
  (\bibinfo {year} {2011})}\BibitemShut {NoStop}%
\end{thebibliography}

\end{document}